\documentclass[slac_one]{revtex4}
\usepackage{graphicx}
\usepackage{fancyhdr}
\begin{document}

\title{Recent HERA Results Sensitive to SUSY} 

%

\author{G. Brandt (for the H1 and ZEUS collaborations)}
\affiliation{DESY, Notkestrasse 85, 22607 Hamburg, Germany}

\begin{abstract}
Recent results of searches for new physics at HERA using
the complete data sample corresponding to about $0.5~$fb$^{-1}$ 
per experiment are reviewed. 
Searches for leptoquarks, isolated leptons (electrons, muons and taus)
and a generic search for new physics in many topologies are reported.
No evidence for new physics is found.
The $ep$ collisions at HERA are ideally suited 
to search for resonantly produced squarks in SUSY with $R$-parity 
violation.
It is highlighted how the present results can be interpreted in this way.
\end{abstract}

\maketitle

\thispagestyle{fancy}


\section{LEPTOQUARKS}

The $ep$ collider HERA located at DESY, Hamburg, 
was ideally suited to search for new particles coupling to 
electron- or positron-quark pairs during its operation time from 1992 to 2007. 
Such leptoquark (LQ) particles could be resonantly produced
up to a centre-of-mass energy of $\sqrt{s} = 320$~GeV. Via contact-interaction-like
exchange of particles the sensitivity of HERA extends to even higher energies.
In the general Buchm\"uller-R\"uckl-Wyler (BRW) framework a total of 14 LQs are
investigated~\cite{Buchmuller:1986zs}.
In SUSY with $R$-parity violation, squarks are 
produced in the same way via Yukawa couplings $\lambda'_{ijk}$,
where the indices denote families~\cite{Butterworth:1992tc,Barbier:2004ez,Aktas:2004ij}.
Two of the LQs, namely the
$\tilde{S}_{1/2}^{L}$ and the $S_{0}^{L}$ in the BRW notation
can be interpreted as $\tilde u_j$ or $\tilde d_k$ squarks
produced via $R$-parity violating
couplings $\lambda^{\prime}_{1j1}$ or $\lambda^{\prime}_{11k}$, respectively.

\begin{figure}
  (a)~\includegraphics[width=0.4\textwidth]{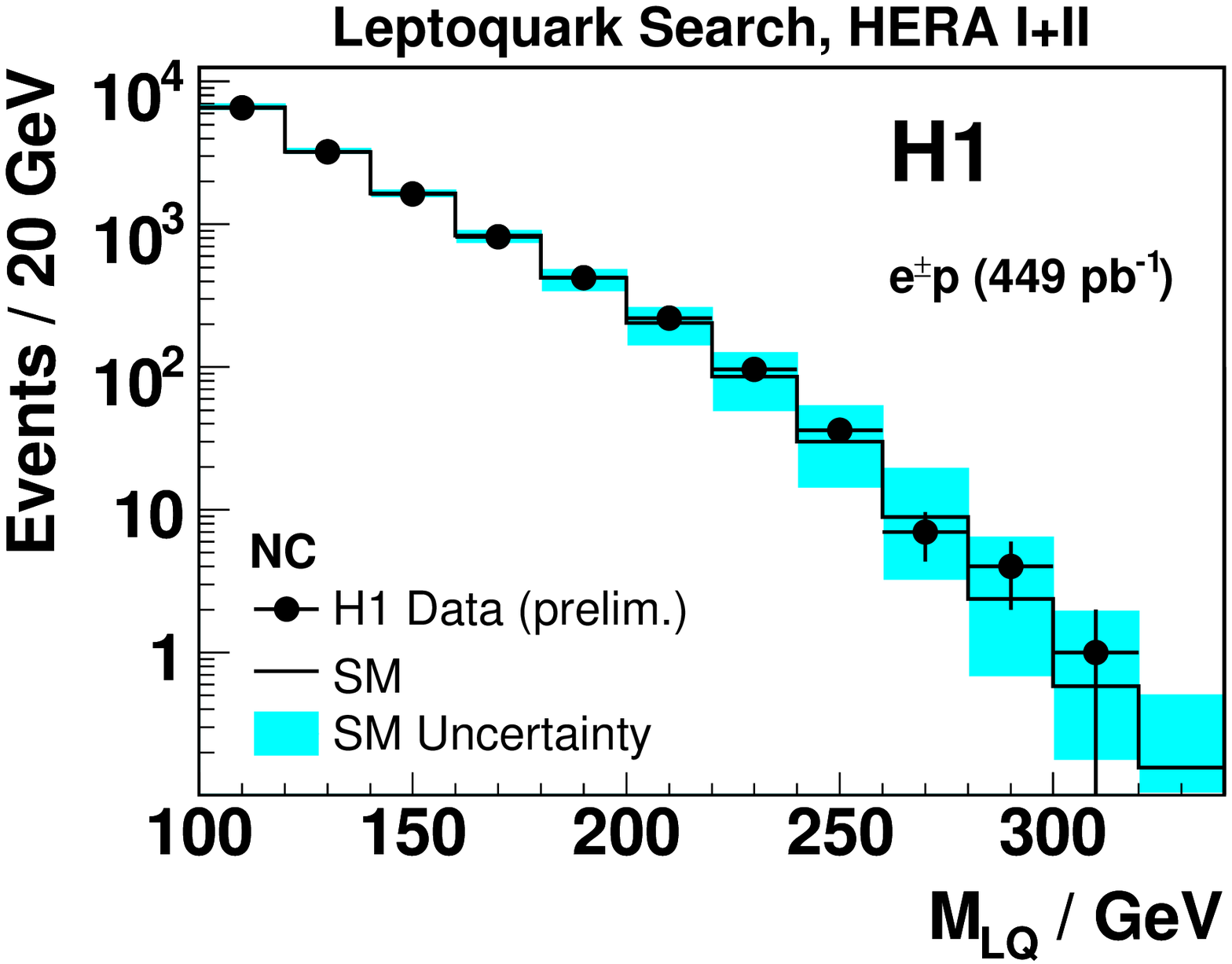}
  (b)~\includegraphics[width=0.4\textwidth]{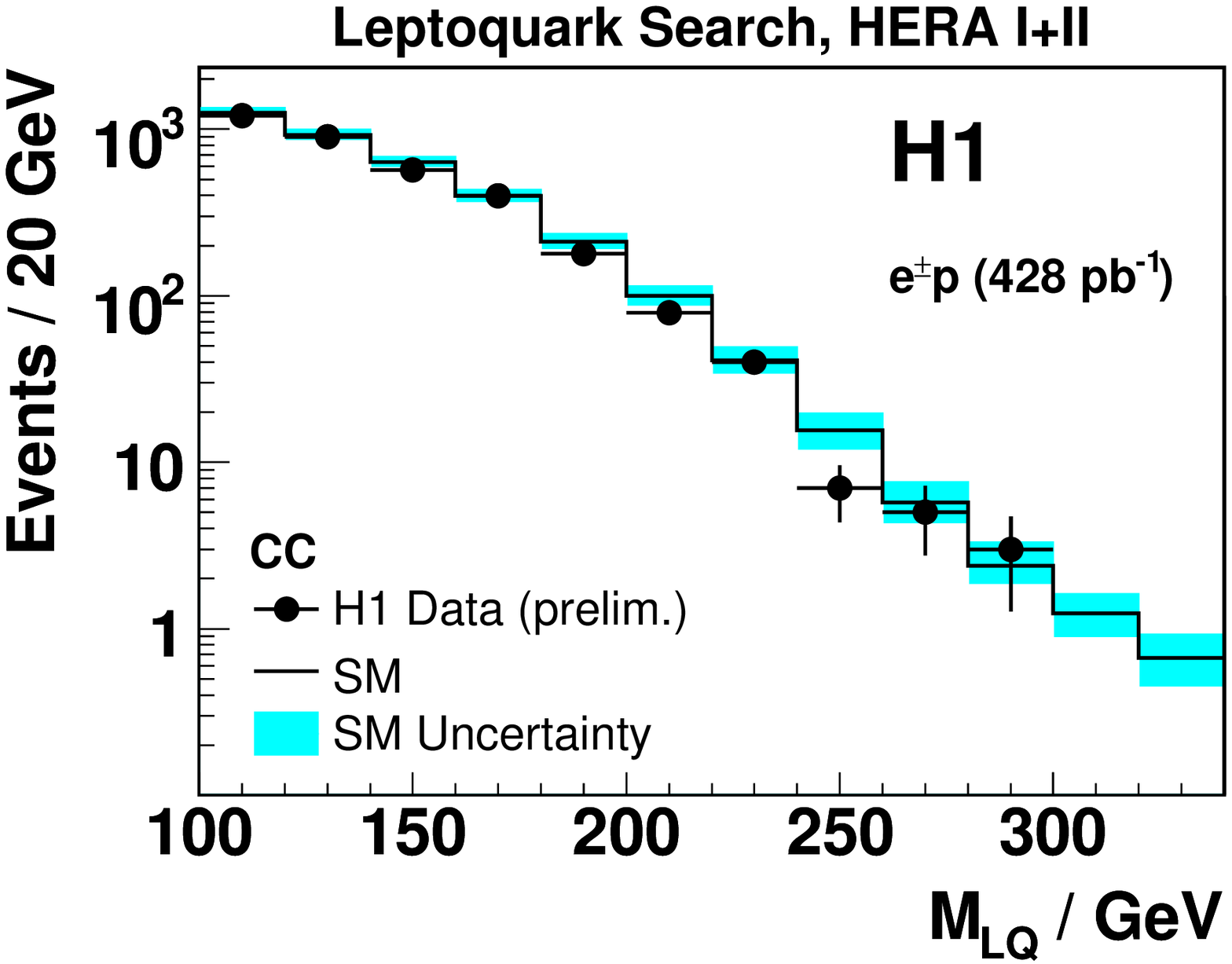}
  \caption{\label{figure:h1lqspectra}Reconstructed mass spectra for the
  HERA~I+II $e^{\pm}p$ data for neutral current~(a) and charged current~(b)
  events in the H1 leptoquark analysis.}
\end{figure}

\begin{figure}
  (a)~\includegraphics[width=0.4\textwidth]{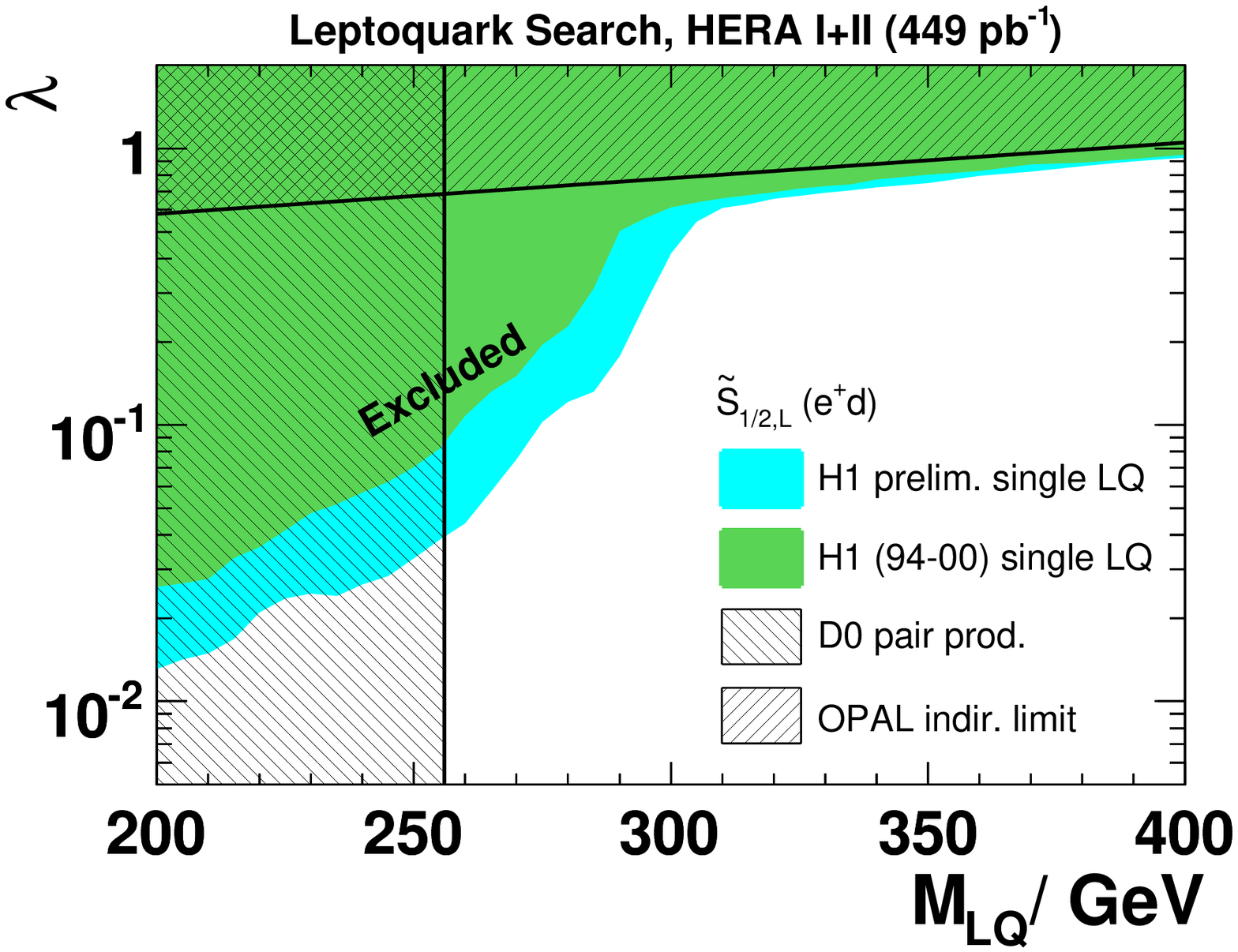}
  (b)~\includegraphics[width=0.4\textwidth]{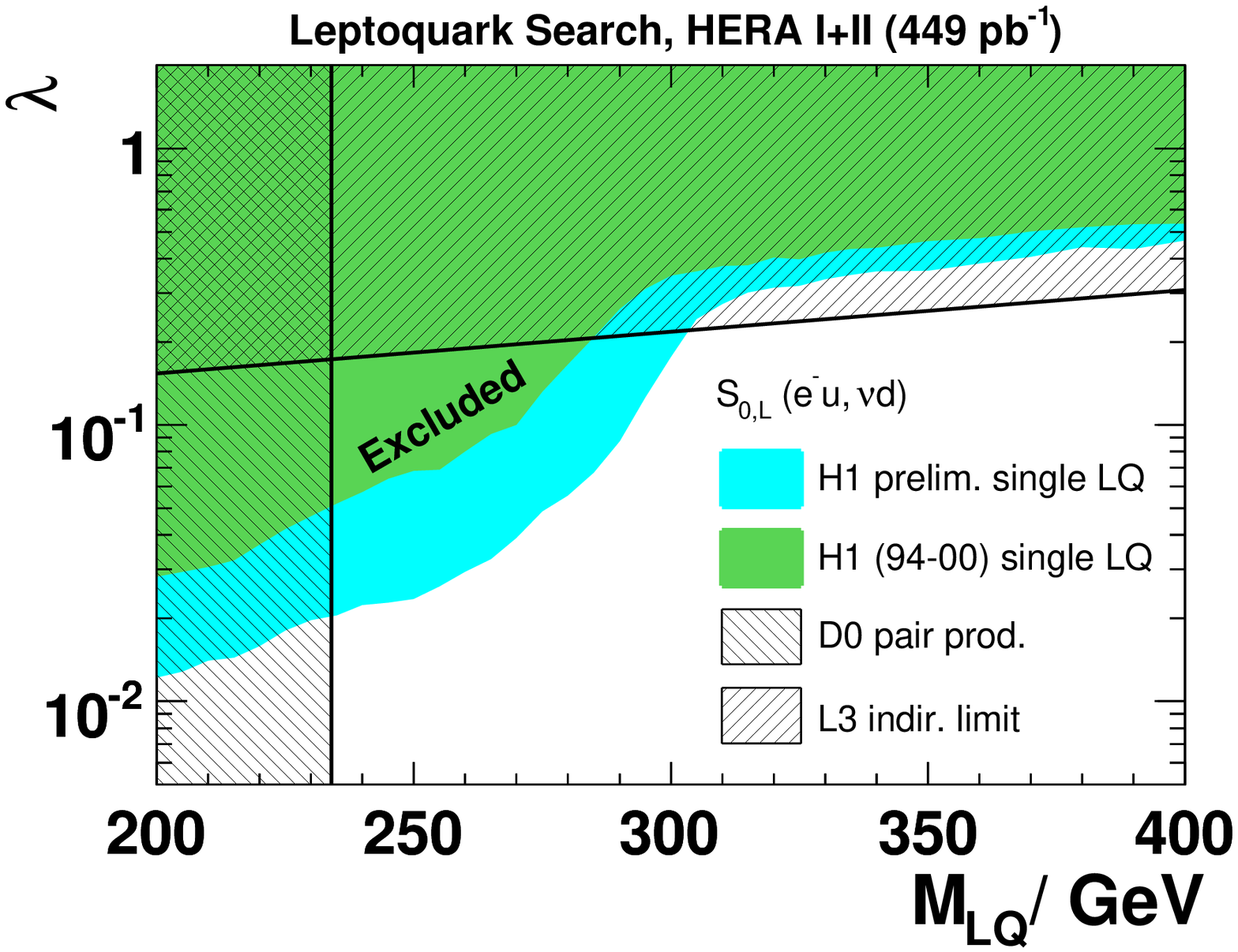}
  \caption{\label{figure:h1lqlimits}H1 preliminary $95\%$ confidence limits
    on the production of $\tilde{S}_{1/2L}$~(a) and $S_{0L}$~(b)
    leptoquarks, corresponding to $\tilde{u}_i$ and $\tilde{d}_j$
    squarks in $R$-parity violating models, respectively. Limits on the coupling
    $\lambda$ are shown as a function of the leptoquark mass.}
\end{figure}

These objects may appear in
the reactions $ep\to eX$ and $ep\to \nu X$, where the $X$ denotes the hadronic
final state. For these reactions there is an irreducible background from 
deep-inelastic neutral and charged current scattering. 
For LQ masses below the kinematic limit of HERA ($320\,\text{GeV}$) one may
see LQs as a resonant structure in the reconstructed mass $M_{LQ}$. 
Figure~\ref{figure:h1lqspectra} shows the
LQ mass spectra seen by the H1 collaboration using the
full HERA dataset~\cite{h1prelimlq}.
No evidence for LQ production is seen, and limits are set on
the LQ coupling $\lambda_{eq}$ as a function of the LQ mass. 
Limits on LQ production at $95\%$ confidence level are shown in Figure
\ref{figure:h1lqlimits}. 
They may also be interpreted as squark limits for 
SUSY models where the direct $R$-parity violating decay dominates or
for models with squark masses larger than the HERA centre-of-mass energy. 
For couplings of electromagnetic strength, $\lambda=0.3$, the
production of $\tilde{S}_{1/2}^{L}$ and $S_{0}^{L}$ LQs are excluded for
masses up to $295$ GeV and $310$ GeV, respectively.

If the LQ mass is higher than the kinematic limit,
LQs may show up as contact interactions. 
For this region, the ZEUS collaboration reports preliminary limits 
on the ratio of LQ mass to coupling
$M/\lambda$, using $274\,\text{pb}^{-1}$ of data~\cite{zeusprelimci}.
For example, the production of $\tilde{S}_{1/2}^{L}$ LQs is excluded for 
$M/\lambda < 0.96$~TeV.

Another model for LQ production at HERA includes flavour-violating
decays. Here the LQ may also have couplings $\lambda_{\mu q'}$ or
$\lambda_{\tau q'}$.
At HERA such models are 
probed in the reactions $ep \to \mu X$ or $ep \to \tau X$ 
\cite{Chekanov:2005au,Aktas:2007ji}.
Again the limits on
the search for $\tilde{S}_{1/2}^{L}$ and $S_{0}^{L}$ LQs may be
interpreted as a search for squarks with off-diagonal $R$-parity violating
couplings $\lambda'_{ijk}$. 
A preliminary search for LQs decaying to
$\mu^{-}+\text{jet}$  has been updated
using the full H1 $e^{-}p$ data~\cite{h1prelimlfv}.
For couplings of electromagnetic strength $\lambda_{\mu
  q'}=\lambda_{eq}=0.3$, assuming $\lambda_{\tau q'}=0$, the production of a
$S_{0L}$ is excluded for masses up to $305$~GeV. 

\section{ISOLATED LEPTONS}

\begin{figure}
  (a)~\includegraphics[width=0.3\textwidth]{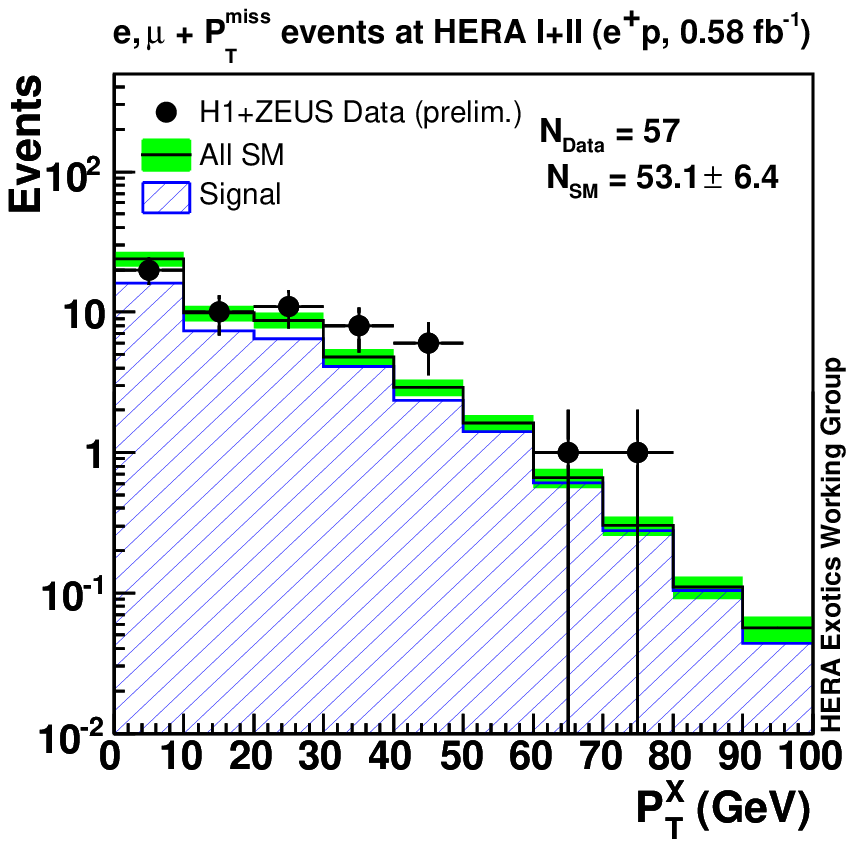}
  (b)~\includegraphics[width=0.3\textwidth]{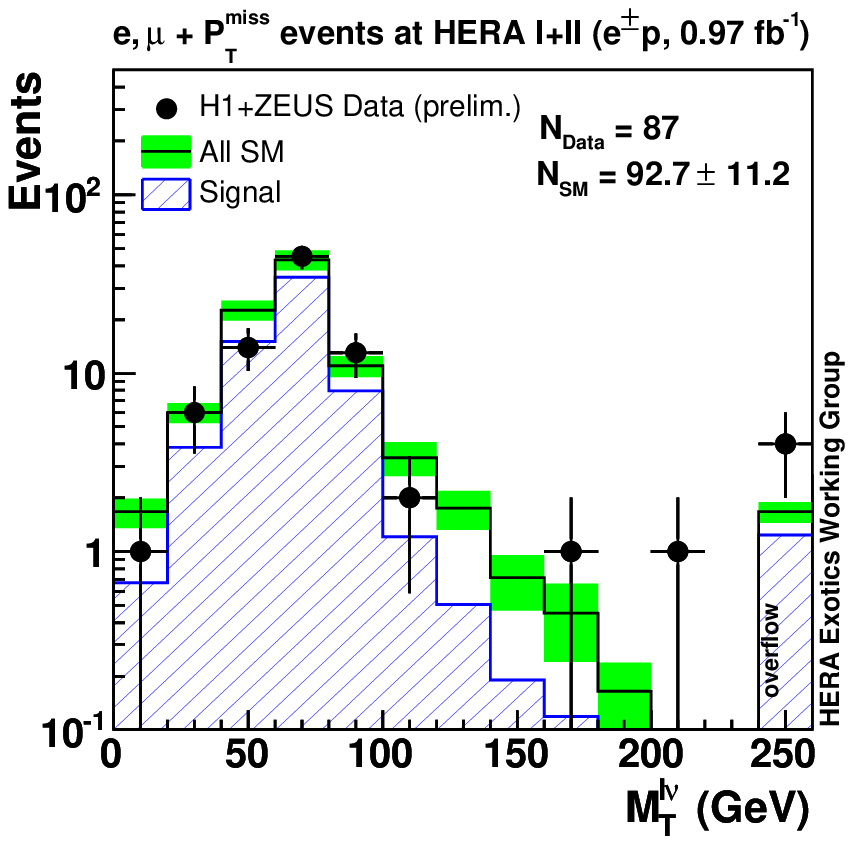}
  \caption{\label{figure:h1isolepptx}\label{figure:h1isolepmt}
    (a)~H1 and ZEUS combined data on 
    isolated leptons ($\ell=e,\mu$) with missing transverse momentum.
    The $e^{+}p$ data are shown as a function of the transverse
    momentum of the hadronic system $P_T^X$.
  (b)~H1 and ZEUS combined data on 
  isolated leptons ($\ell=e,\mu$) with missing transverse momentum.
  The data are shown as a function of the transverse mass.}
\end{figure}

The HERA data are searched for events with an isolated lepton 
($\ell=e,\mu$) with high transverse momentum
$P_T^{\ell}>10\,\text{GeV}$ and high missing transverse momentum
$P_T^{\text{miss}}>12\,\text{GeV}$. The $P_T^{\text{miss}}$
is attributed to a neutrino which escaped detection. 
SUSY models predict an enhanced rate of such events with isolated leptons 
in $ep$ collisions, for example due to resonant $\tilde t$ production, 
as has been investigated in HERA-I data~\cite{Aktas:2004tm}.
Such heavy resonances 
would be expected to produce an excess over the SM predictions at
large values of total hadronic transverse momentum $P_T^X$.
The data from both experiments, ZEUS and H1, are combined in a common
phase-space~\cite{h1zeusprelimisolep}. 
Figure~\ref{figure:h1isolepptx}~(a) shows the distribution of
$P_T^X$ for the $e^{+}p$ data. At high $P_T^X$, where the SM
prediction is small, there is an excess of
events, with $23$ observed over $14.6\pm 1.9$ expected from Standard
Model (SM) processes. No such excess is present in the $e^{-}p$ data.
The transverse mass of the lepton and the neutrino for the complete $e^{\pm}p$ data 
is shown in Figure~\ref{figure:h1isolepmt}~(b). There is clear evidence that the events are
dominated by single $W$ production. 

An enhanced production of real $W$ bosons with high $P_{T}^{X}$ at HERA
may also originate from the decay of top quarks, 
as previously investigated with HERA~I data~\cite{Chekanov:2003yt,Aktas:2003yd}.
While the SM predicts a cross section too small to observe in the available
data, many models, including $R$-parity violating SUSY~\cite{Yang:1997dk}, 
predict an observable rate of anomalously produced top quarks.
The H1 collaboration reports a
preliminary $95\%$ confidence limit on the anomalous coupling
${\kappa}_{tu\gamma}<0.14$, using their complete dataset \cite{h1prelimtop}.

Also interesting in the context of SUSY is the production of tau leptons, which
exhibits an  enhanced rate over SM expectations in many SUSY scenarios, for 
example if the  LSP is $\tilde\tau_1$ decaying via 
$\tilde\tau_1 \rightarrow \tau \nu_i$.
H1 has searched for events with isolated tau leptons and 
missing transverse momentum $P_{T}^{\rm miss}>12$~GeV, where the tau lepton 
decays hadronically~\cite{h1prelimtau}. 
The tau is identified by looking for a narrow jet of $P_{T}>7$~GeV with 
exactly one isolated track (``1-prong'' signature). 20 events are observed in 
the data, which agrees well with the SM expectation  of $19.5\pm3.2$. 
This expectation is dominated by irreducible CC background.

\section{GENERAL SEARCH}

\begin{figure}
  \begin{center}
    (a)~\includegraphics[width=.48\textwidth,angle=-90]{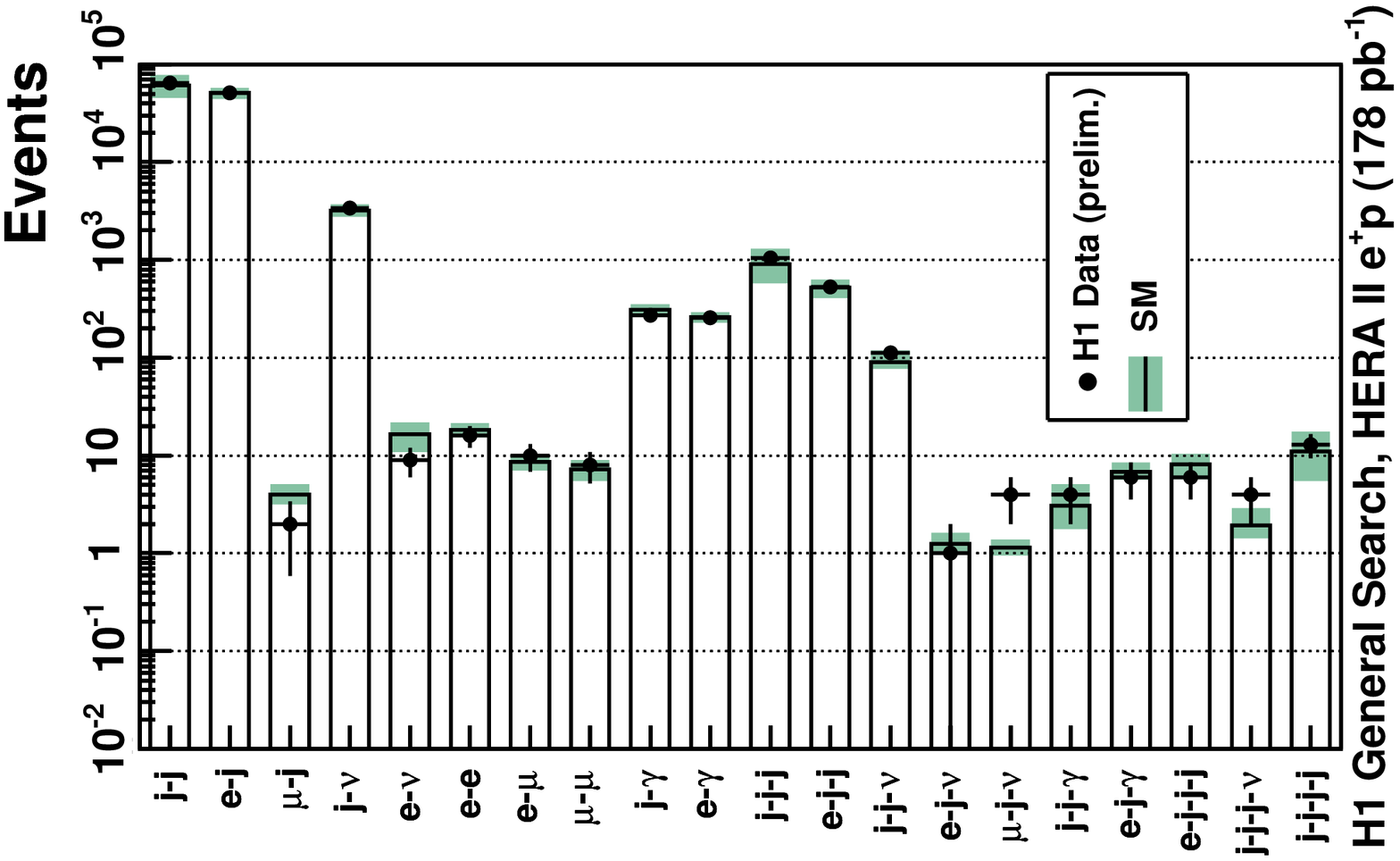}
    (b)~\includegraphics[width=.48\textwidth,angle=-90]{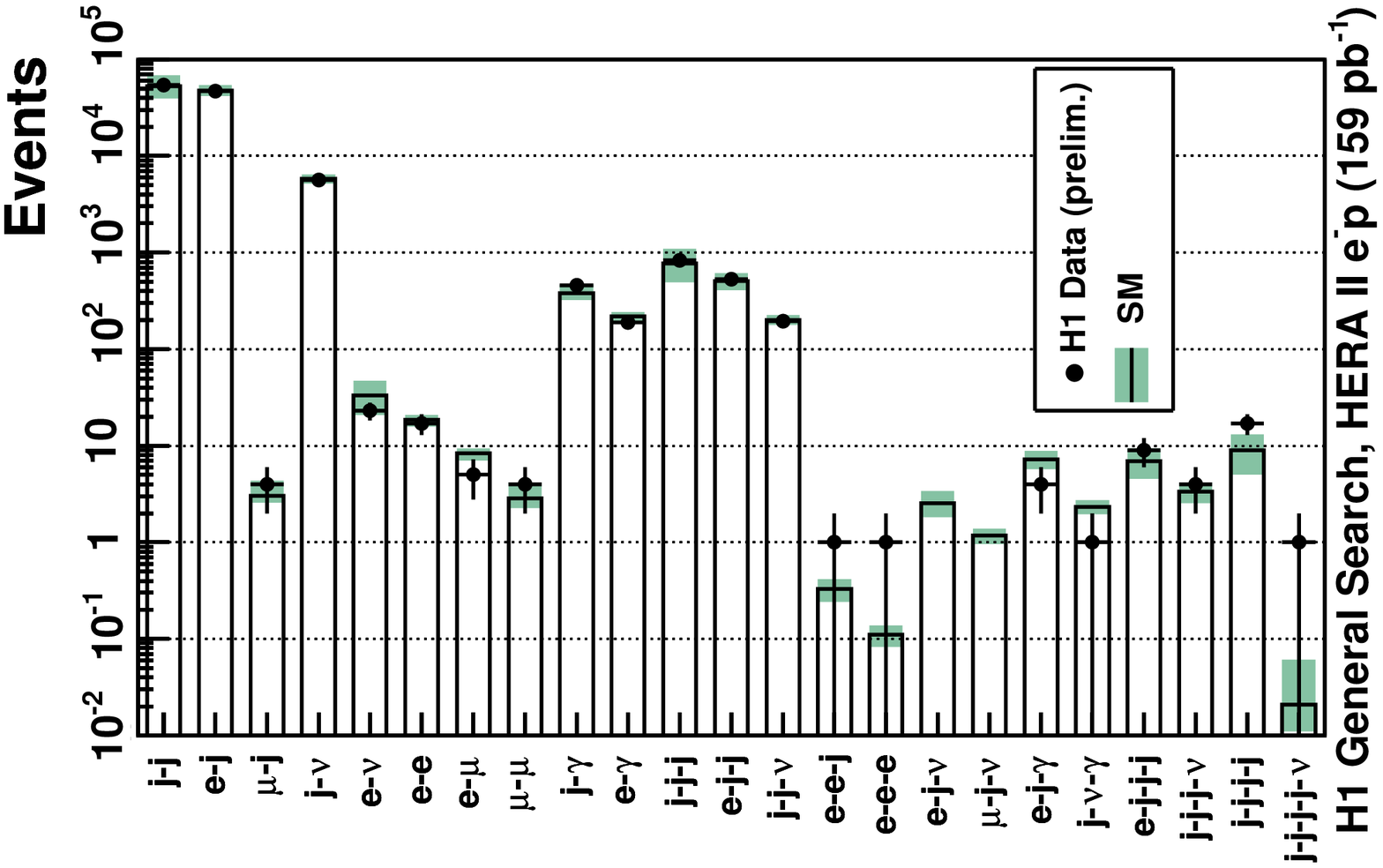}
  \end{center}
  \caption{The data and the SM expectation in event classes investigated by the H1 
  general search. Only channels with observed data events or a SM expectation greater 
  than one event are displayed. The results are presented separately for $e^+p$ (a) and
   $e^-p$ (b) collision modes.}
\label{fig:GS}  
\end{figure}

From previous searches for $R$-parity violating SUSY at HERA it is known
that the inclusion of as many final state topologies as possible is necessary
to be sensitive to SUSY fairly independently of the model parameters~\cite{Aktas:2004ij}.
%
%
In a general search developed by the H1 Collaboration on HERA~I
data~\cite{Aktas:2004pz} all final states containing at least two objects 
($e$, $\mu$, $j$, $\gamma$, $\nu$) with 
$P_T >$~$20$~GeV in the polar angle range  $10^\circ < \theta < 140^\circ$ are 
investigated~\cite{h1prelimgs}.
The observed and predicted event yields in each channel are presented in 
Figure~\ref{fig:GS}(a) and (b) for $e^+p$ and $e^-p$ collisions, respectively.
While the interesting events observed in the search for isolated leptons
are found again, overall good agreement between data and SM prediction is 
observed.

\section{SUMMARY}

New results from searches for leptoquarks, isolated leptons with
missing transverse momentum and a general search at HERA have been
reported. No evidence of new physics has been found, but a good
agreement of data and SM expectations is found in all channels.
While SUSY in the reach of HERA can be excluded with these
results to the extent discussed in these proceedings, a dedicated interpretation 
of the HERA data in terms of $R$-parity violating SUSY by means of an extensive 
paramater scan of the phase space to exclude SUSY is still possible.

\end{document}